\documentclass{elsart}
\usepackage{epsfig}
\usepackage{amssymb}

\begin{document}
\begin{frontmatter}

\title{Towards a new concept of photomultiplier based on silicon technology} 

\author[a,b]{S.Privitera}
\author[a,b]{S. Tudisco}
\author[a,b]{L.Lanzan\'o}
\author[a,b]{F.Musumeci}
\author[a]{A.Pluchino}
\author[a,b]{A.Scordino}
\author[a]{A.Campisi}
\author[a]{L.Cosentino}
\author[a]{P.Finocchiaro}
\author[c]{G.Condorelli}
\author[c]{M.Mazzillo}
\author[d]{S.Lombardo}
\author[d]{E.Sciacca}

\address[a]{INFN-LNS, Via Santa Sofia 65 95125 Catania Italy}
\address[b]{DMFCI and Dipartimento di Fisica ed Astronomia, Università di Catania, Viale A. Doria 6, 95125 Catania, Italy}
\address[c]{ST-Microelectronics, Stradale Primosole 50,  95100 Catania Italy}
\address[d]{IMM-CNR, Stradale Primosole 50,  95100 Catania Italy}

%

\begin{abstract}


In order to build a new concept of photomultiplier based on silicon technology, design and characterization of 5x5 arrays of a new generation of single photon avalanche diodes (SPAD) manufactured by ST-Microelectronics have been performed. Single photons sensitivity, dark noise and timing resolution of the SPAD-STM devices in several experimental conditions have been evaluated. Moreover, the effects arising from the multiple integration of many elements and the study of their common read-out have been deeply investigated.

\end{abstract}

\begin{keyword}
Single photon, avalanche photodiode, silicon photomultiplier, quantum efficiency, picosecond timing, afterpulsing, cross-talk, 2D array.
\end{keyword}
\end{frontmatter}

\section{Introduction}

In the last three decades several groups investigated the possibility to build a silicon photo-sensor suitable for single photon counting applications (\cite{cova1} and references there in). The original idea firstly proposed by R.J. Mc Intyre \cite{mcintyre1} was to
implement a semiconductor photodiode with characteristics suitable for the triggered avalanche operation mode and then able to detects single photons \cite{mcintyre2,goetzberger,haitz1,haitz2} (whence the name Single Photon Avalanche Diode - SPAD). When a p-n junction is reversely biased 10-20 \% above the breakdown voltage value, a single charge carrier entering inside the high field region of the depleted volume can trigger the avalanche multiplication process. The fast leading edge (rise time less then 1 ns) of the corresponding current pulse can be used for detecting and timing the single photo-generated carrier. In that condition, due to the “detectable” value of the flowing current, single optical photons can be detected.
The diode current is negligible until the first carrier generated in the junction depletion layer, impact, ionizes, thus triggering a diverging avalanche process. A suitable circuit, usually called quenching circuit (passive or active), senses the rise of the diode current and quenches the multiplication process by lowering the bias voltage down below the breakdown. 
To be used as SPAD, a diode must have a structure that fulfills some basic requirements: 
(i) the breakdown must be uniform over the whole active area in order to produce a standard macroscopic current pulse; (ii) the dark counting rate must be sufficiently low;
(iii) ) the probability to generate after-pulses should be low;
In dark condition, the carrier sources are essentially two, the diffusion current by quasi neutral regions, which is normally negligible \cite{sciacca1}, and the generation of electrons or holes from trap levels located in the depletion layer. In order to satisfies  (ii) and (iii) prescriptions, both the effects of thermal carrier generation and trapping should be minimized.
The fully characterization of the SPAD device, for single photons detection and timing applications require the estimation of some important figures of merit: the dark counting rate (thermal and afterpulsing effects), the photon detection efficiency, the time resolution, the maximum capable excess bias voltage, the optimal temperature condition, the hold-off time.
\\
Today modern technology gives also the possibility to produce SPAD detectors with an integrate quenching mechanism based on a Metal-Resistor-Semiconductor structure. Precise resistive elements are embedded for each individual microcell and provides effective feedback for stabilization and quenching of the avalanche process \cite{saveliev}. Such technology allows the production of large numbers of microcells in very fine structure on a common substrate with common electrodes, to provide a proportional mode operation device (Silicon Photomultiplier SiPM) able to detect and count photons \cite{buzhan}. 
In this paper we present the fully characterization of the performances of new generation devices manufactured by ST-Microelectronics in Catania.

\begin{figure}
\begin{center}
\epsfig{figure=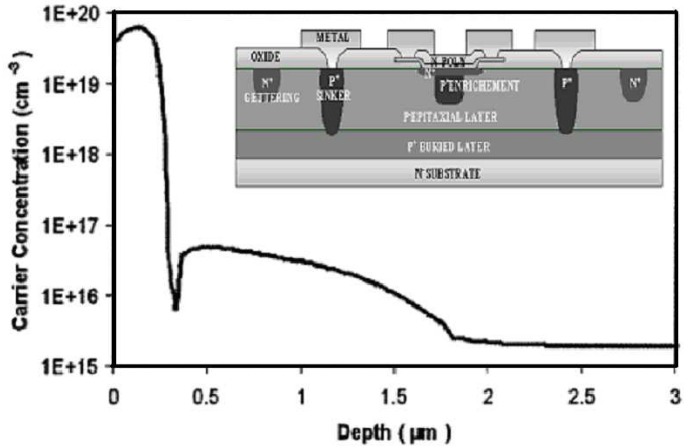,width=10cm,angle=0}
\end{center}
\caption{Vertical cross section of the SPAD device and the profile of carrier concentration inside the junction.
}
\end{figure}

\section{The fabrication process }

Figure 1 shows the cross section of the SPAD structure. The process starts with a Si <100> n-substrate on which is grown a boron doped epitaxial layer with a p+ buried layer and with a p- doped layer. The reason to form a buried p-n junction is twofold. First, the detector time response is improved because the effect of photo-generated carriers diffusing in the undepleted region is reduced \cite{ripamonti}. Second, isolation with the substrate is introduced and makes possible the monolithic integration of various SPADs and other devices and circuits. The p+ buried layer is necessary to reduce the series resistance of the device. The p- layer must be thin enough to limit the photo-carrier diffusion effect above mentioned. A good tradeoff has to be found for this thickness, because if it is made too thin the edge breakdown occurs at a voltage not much higher than the breakdown voltage of the active area. In order to reduce the contact resistance of the anode and provide a low resistance path to the avalanche current, the p+ sinkers are then created with a high-dose boron implantation step. 
\\
The next step, consisting on a local gettering process, is a key step in the process and was introduced in the last recipe. At this point of the process a heavy POCl3 diffusion through an oxide mask is made on the topside of the wafer close to the device active area. Heavy phosphorus diffusions are well known to be responsible for transition metal gettering \cite{heislmair}. Unfortunately, the well-known phosphorous pre-deposition on the backside of the wafer is not able to getter the distant active area of the device because metal (Pt, Au , Ti) too slow during the final anneal. For this reason, if the gettering sites are created suitably close to the active region, a major improvement is observed.
The next step is the p+ enrichment diffusion obtained with a low energy boron implantation, producing a peak concentration of 5x1016 cm-3 , followed by a high temperature anneal and drive in \cite{lacaita1}.  
The first generation of devices was fabricated with a deposited polysilicon cathode doped by Arsenic implantation and diffusion. In order to damage as little as possible the active area of the device, the As+ ion implantation energy was carefully calculated; nevertheless, devices with very high dark-counting rate have been obtained. A remarkable improvement was obtained in the second generation by doping in situ the polysilicon. Further improvement was achieved in the third generation by accurately designing a Rapid Thermal Anneal to create a precisely controlled shallow Arsenic diffusion below the polysilicon in the p-epilayer. The final net doping profile has been measured by spreading profiling and it is shown in figure 1. An important issue for the high SPAD quality is the uniformity of the electric field over the whole active area. If the electric field is not uniform, the quantum detection efficiency (QE) of the device becomes dependent on the absorption position on the active area. The lower the electric field the lower the QE. Quality of the manufactured photodetectors has been checked by means of Emission Microscopy measurement \cite{sciacca2}. 
\\
\begin{figure}
\begin{center}
\epsfig{figure=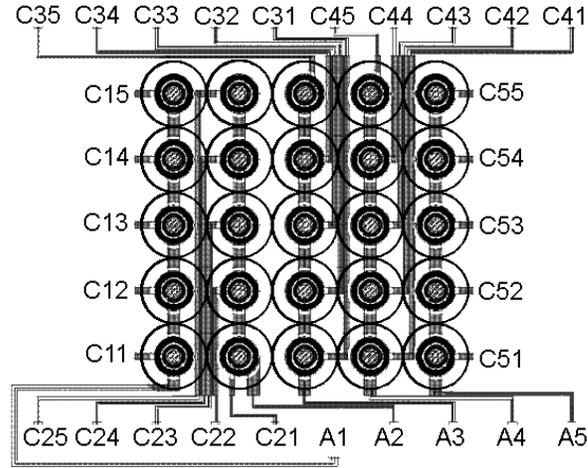,width=9cm,angle=0}
\end{center}
\caption{Lay-out of the 2-D array of 5x5 SPAD devices with the particular active area of 20 $\mu$m of diameter.
 }
\end{figure}
Planar view of an array manufactured by integration of 25 pixels in square geometry 5X5 is reported in figure 2. Pixels of three different diameter dimensions have been chosen for the integration: 20, 40 and 60 $\mu$m. Separation distances between adjacent pixels, according to different diameters, ranges between 160 $\mu$m and 240 $\mu$m. Anode contacts are in common for each row, while each cathode is separately contacted and available to the external by different pads. One difference concerning the vertical structure of the array, respect to the single pixel construction, is the gettering region which uniformly surrounding the active area of the pixel. Breakdown voltages distribution over an array has a mean value of 31V for the 20 m device with a standard deviation of  0.08 V. In order to study the reduction of the optical cross-talk contribution, arrays that are optically and electrically isolated by deep thin trench technology have been already designed and fabricated. The trench process starts with a vertical etch 10 m deep and 1m large, a subsequent oxide deposition for complete electrical isolation. The process continues with tungsten fill to avoid optical crosstalk and ends with planarisation. As it will be shown from measurements reported in the next sections, the lower average dark count rates indicate that trenches processing had the role of an efficient gettering function.

\section{SPAD operating conditions}

Bias supply voltage exceeds breakdown voltage of the junction by an amount called excess bias voltage (EBV) or over-voltage, which has fundamental influence on the detector performance. A photon is detected if it is absorbed inside the sensitive volume and, due to the very high field, if the primary generated carriers trigger the avalanche multiplication process; since a higher electric field enhances the probability to trigger the avalanche. Photon detection efficiency PDE then increases with the over-voltage, it was measured at room temperature, for 20\% of EBV it was about: 50\% at 550 nm, 10\% at 850 nm and 3\% at 1000 nm \cite{belluso}.  
\\
As previously mentioned, lowering the bias voltage to breakdown or below its, the current is quenched and the operating bias voltage is restored in order to be able to detect another photon. This operation requires a suitable circuit that must sense the leading edge of the avalanche current, generate a standard output pulse that is well synchronized to the avalanche rise, quench the avalanche by lowering the bias below the breakdown voltage and restore the photodiode voltage to the operating level. The features of the quenching circuit dramatically affect the operating conditions of the detector and, therefore, its actual performances.
\\
The basic idea of active quenching circuit AQC was simply to sense the rise of the avalanche pulse and react back on the SPAD, forcing, with a controlled bias-voltage source, the quenching and reset transitions in short times. The rise of the avalanche pulse is normally sensed by a fast comparator whose output switches the bias voltage source to breakdown voltage or below. After an accurately controlled hold-off time, the bias voltage is switched back to operating level. A standard pulse synchronous to the avalanche rise is derived from the comparator output to be employed for photon counting. The basic advantages offered by AQC approach are the fast transition and the short and well defined duration of the avalanche current and of the dead time \cite{tudisco1}.
\\
The Passive-Quenching Circuit, PQC, was widely discussed \cite{tudisco2}; the avalanche current is quenched itself by increasing a voltage drop on a high impedance load. The SPAD is reverse biased trough a high ballast resistor $R_B$ of 100 K$\Omega$ or more, the junction capacitance value is typically a few hundred of fF, and stray capacitance (to ground of the diode terminal connected to $R_B$, typically a few pF. The diode resistance depends on the semiconductor device structure, and is of the order of  some hundred of $\Omega$s. Avalanche triggering corresponds to closing the switch in the diode equivalent circuit. The avalanche current discharges the capacitance so that diode voltage and diode current exponentially fall \cite{tudisco1}[15]. Avalanche quenching corresponds to the opening of switch in the diode equivalent circuit. Small current in ballast resistor $R_L$ slowly recharges the capacitances; the diode voltage exponentially recovers toward the bias voltage. A photon that arrives during the first part of the recovery is almost certainly lost, since the avalanche triggering probability is very low. The output pulse from a PQC can be obtained by inserting a low value resistor $R_L$ in series (50 $\Omega$) on the ground lead of the circuit \cite{tudisco1}[15]; in a such contition the pulse waveform is directly determined by the diode current. The ST-technology gives also the possibility to produce SPADs with an integrate quenching resistor $R_B$ and then the possibility to realize the so-called Silicon-Photomultiplier; in that direction is devoted the complete characterization of the device here presented.  
\begin{figure}
\begin{center}
\epsfig{figure=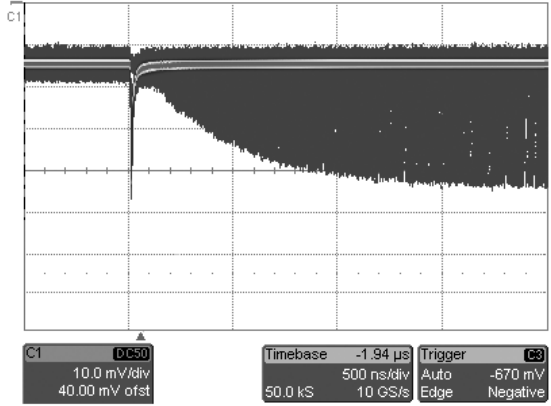,width=9cm,angle=0}
\end{center}
\caption{Persistence of signals on the digital oscilloscope: electrical pulses arise from a 20 m pixel passively quenched SPAD-STM, at the temperature of 25 °C, with a 100 k$\Omega$ ballast resistor RB, biased at about 10\% of EBV. 
 }
\end{figure}

\section{General features}

A typical (anodic) signal from the passively quenched SPAD-STM devices is characterized by a very fast climb up to a maximum positive value follows by a slow tail. The fast rising time of the SPAD pulse, about few hundreds of picoseconds, is due to the avalanche formation, so connected to the intrinsic characteristics of the diode. On the other hand, the successive fall time is drove by the quenching circuit, its trend is exponential and the time constant  \cite{cova2} was estimated of the order of 30 ns. The maximum amplitude of the signal depends on both the series resistor RL and the value of EBV; typical value is about 40 mV at 10\% of EBV.  
After the primary avalanche the pulse maximum amplitude is exponentially recovered towards the original operating condition with a time of about 1.5 $\mu$s (see figure 3). This value remains approximately the same varying both the temperature and the excess bias voltage and may affect the performances of the device only in those applications which require very high counting rate.
\\
\begin{figure}
\begin{center}
\epsfig{figure=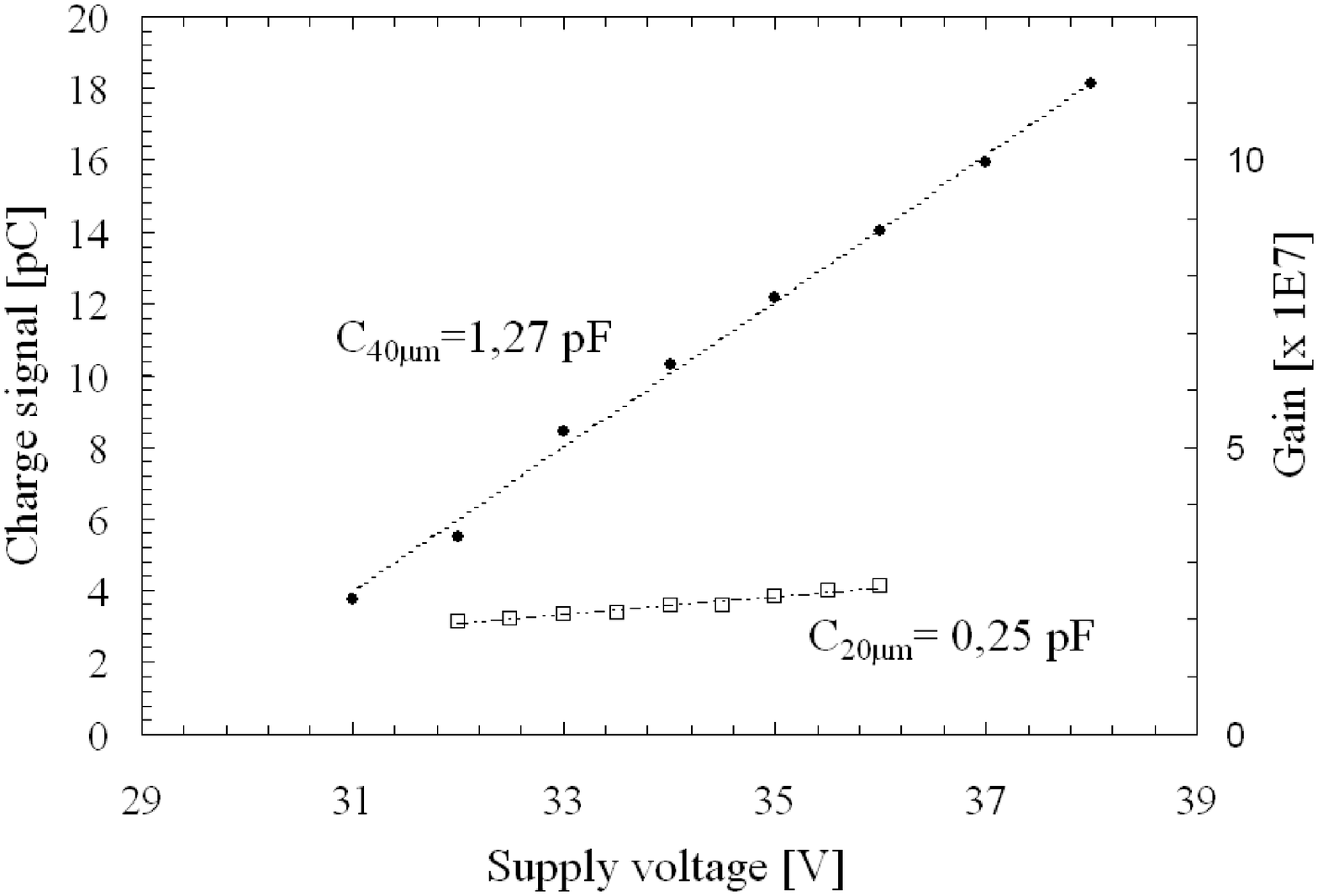,width=12cm,angle=0}
\end{center}
\caption{Capacitance values and the calculated gain for devices of 20 and 40 $\mu$m on active diameters . 
 }
\end{figure}
Total device capacitance and gain values were extrapolated by the linear fit relating the charge accumulated and measured during the flowing of the avalanche current as a function of the bias voltage, for both 20 m and 40 m devices, as is presented in figure 4; the respective values of 1.27 pF and 0.25 pF and their related gains were extracted. 
Another peculiarity arising from the observation of the pulse profile was the occurrence of a non adequate quiescent condition of the device, signed by the non continuous decay of the signal to the ground level, as showed by the arrow in figure 5. The timing duration of such quiescent state is a function of the EBV and is related to residual current determined by the lowest value of the ballast resistor RB \cite{cova2}. Such phenomenon affects the trend of the dark counting rate as a function of EBV and, in particular, was related to its reduction over a certain limit value of excess EBV, see figure 6.
It’s well known that, also in absence of illumination, thermal generation effects produce current pulses which represent the internal noise of the detector. As previously mentioned, the detector noise is due to both i) the Poissonian contribution of the dark counts arising by the thermal carriers generation, and ii) the occurrence of delayed pulses, due to the trapping and the delayed releasing of the avalanche carriers from deep levels inside the junction. Those released carriers have a certain probability to trigger other avalanches; the so-called afterpulsing effect may affects the photon counting.
\\
\begin{figure}
\begin{center}
\epsfig{figure=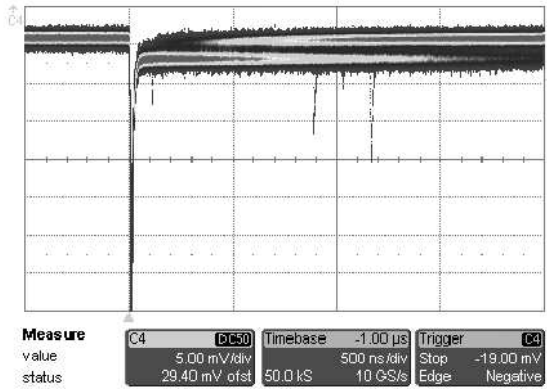,width=8cm,angle=0}
\end{center}
\caption{Persistence of signals on the digital oscilloscope: electrical pulses arise from the 20 $\mu$m pixel passively quenched SPAD-STM, at room temperature of 25 °C, biased at about 10\% of EBV. The non continuous decay of the signal to the ground level is evident. 
 }
\end{figure}
The dark counting rate increases with the EBV because of two effects: i) the field-assisted enhancement of the emission rate from generation centers and ii) the increase of the avalanche triggering probability \cite{privitera}. In respect of the temperature, by using a dedicated cooling system, which stabilizes the temperature of the package where the sample detector is mounted, a set of measurements of dark counting rate (pixel by pixel) operated with a scaler, as a function of both the EVB and the temperature, have been performed and reported for both the 20 and 40 $\mu$m devices in figure 6. Typical values of dark counting rate at room temperature and at 10-15\% of EBV was calculated: 400 cps for the 20 $\mu$m and 2000 cps for the 40 $\mu$m pixel. We observed the expected enhancement of the dark counts with both the temperature and the detector dimensions; the increasing with the temperature is connected to the intrinsic nature of the dark counts.
On the other side, the subsequent decreasing with the bias voltage is expected because the occurrence of the residual charge effect, discussed in the previous section; the phenomena rightly observed for values of the excess bias voltage greater than a reference value, when the delayed restoring of the operating condition of the device overcomes the recharging time, determining a reduction of the counting rate. 
It is worthy that, in order to deeply evaluate the effect of temperature on dark counting rate, is necessary to well investigate the afterpulsing contribution.
\begin{figure}
\begin{center}
\epsfig{figure=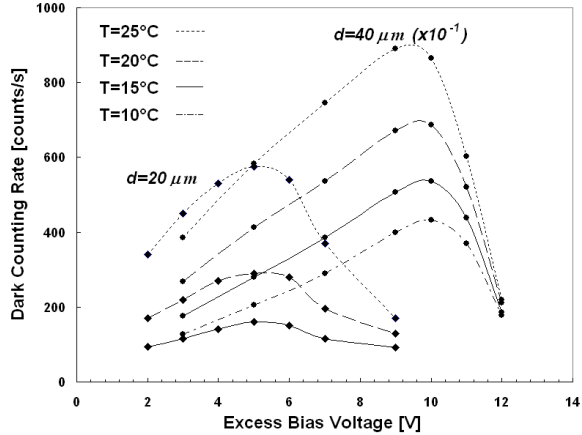,width=11cm,angle=0}
\end{center}
\caption{Dark counting rate as a  function of the excess bias voltage, measured for passively quenched SPAD devices of active area with diameters d=20 $\mu$m (full diamonds) or d=40 $\mu$m (full circles), at different temperatures: (dotted) T=25°, (dashed) T=20°, (solid) T=15°, (dash-dotted) T=10°. In order to plot both the set of data, counts arising from the bigger device was reduced a factor 10.
 }
\end{figure}

\subsection{The afterpulsing phenomenon}

During the avalanche process some carriers may be captured by deep levels of the depletion region and subsequently released with a statistically fluctuating delay, whose mean value depends on the levels actually involved \cite{privitera}. Released carriers may retrigger the avalanche and generate after pulses correlated with a previous one. The number of trapped carriers during the avalanche increases with the total number of carriers crossing the junction and then with the avalanche current; thus, these so-called afterpulses increase with both the delay of the avalanche quenching and the current intensity. Especially in the passive quenching strategy, the avalanche current is proportional to the EBV, which is chosen in order to perform the best operative conditions in terms of photon detection efficiency and/or of timing  performance \cite{belluso,tudisco1}. An alternative method to minimize the number of trapped charges per pulse requires a dedicated active circuitry, which acts on the quenching delay and reduce the current flowing across the junction.
\\
\begin{figure}
\begin{center}
\epsfig{figure=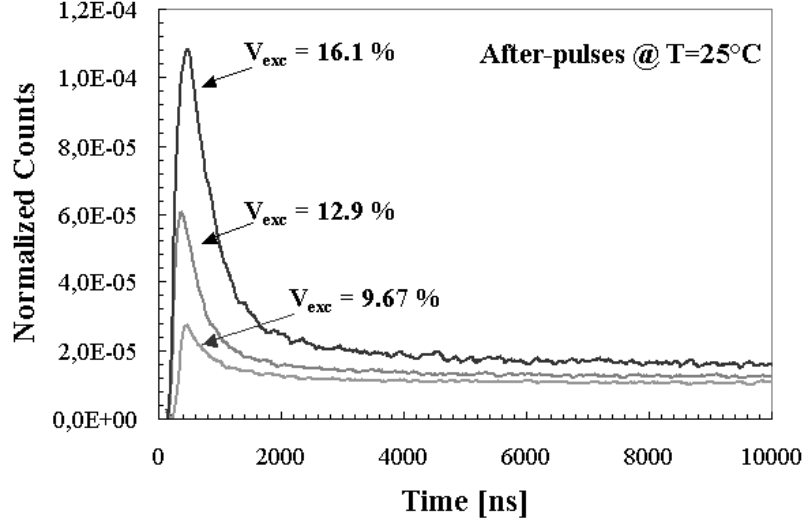,width=11cm,angle=0}
\end{center}
\caption{ Timing distribution of the start-correlated events, for the three particular values of excess bias voltage and at room temperature. Counts were normalized to the total number of triggers.
 }
\end{figure}
A suitable technology must reduce both the generation and recombination centers to a very low concentration level and minimize the concentration of trapping levels. An appreciable improvement on the presented SPAD device has been observed with (i) the substitution of the in situ n-doped polysilicon layer to the implanted one, (ii) a local gettering process. Due to the uniform defect concentration over the device volume, a linear trend of the enhancement of dark counting rate versus active area has been measured in tests carried out on 5x5 arrays \cite{sciacca3}. 
An evaluation of the afterpulses distribution on passively quenched 20 $\mu$m devices, have been performed by means of  a 32 channel multi-hit Time to Digital Converter (TDC), mounted on VME bus and part of a data acquisition system realized with standard nuclear electronics. Such TDC module is able to collect successive events in a 50 $\mu$s, with a sensitivity of 100 ps. The whole system was arranged in order to collect the after-pulses succeeding a reference trigger signal, given in our case by a dark event \cite{privitera}.
\\ 
Experimental set-up and measurements have been realized in order to investigate temperature and EBV dependences on the after pulses. Some results of the investigation are presented in the figure 7: timing distributions of the after pulses succeeding a primary avalanche (the trigger), normalized to the total number of triggers, for the particular cases of three values of EBV and at a fixed temperature have been reported. 
Such distributions was characterized by both the two contributions: the correlated start events, representing the effective after pulses distribution and the uncorrelated background representing the pure thermal dark counting rate of the detector. It is also important to observe the suppression of events in the first hundreds of nanoseconds, due to the quenching and the successive recharging phase joined  to the physical cut-off of the discriminator threshold. 
Comparison between dark count probability and total afterpulsing probability, after the subtraction of the uncorrelated background, had been evaluated by integrating events in a 10 s window; results are reported on table 1 and figure 8. As expected, it should be observed a temperature dependence. The particular: steeper decreasing on the contribution of thermal dark counts respect to the afterpulsing .
In order to perform a deeper investigation on the nature of the afterpulsing phenomenon and then on the involved trap levels a more complex analysis, often used to characterise the timing relaxation of other complex systems \cite{austin}, was requested. We start from the reasonable assumption that the number of after pulses, n(t), represents a convolution of single exponential decay of the type:
\begin{equation}
    n(t)= \int_0^{infty}  N P(\gamma) e^{-\gamma t} dt
\end{equation}
where $\gamma$ denotes the rate constant of a process decay, N a normalization factor, and the probability density function p($\gamma$) represents the occurrence probability of an exponential decay characterised by a lifetime $\tau$ = $\gamma^{-1}$.  According to Eq.(1), n(t) is the Laplace’s transform of the function p($\gamma$), that is n(t) = $\Lambda${p($\gamma$)}.
Starting from the experimental data n(t), the probability density function p($\gamma$) can be obtained by the inverse Laplace’s transform of n(t): p($\gamma$)= $\lambda^{-1}${I(t)}. Such inverse procedure can be analytically performed if data may be reproduced by a Laplace-inverted function, here n(t). 
\begin{figure}
\begin{center}
\epsfig{figure=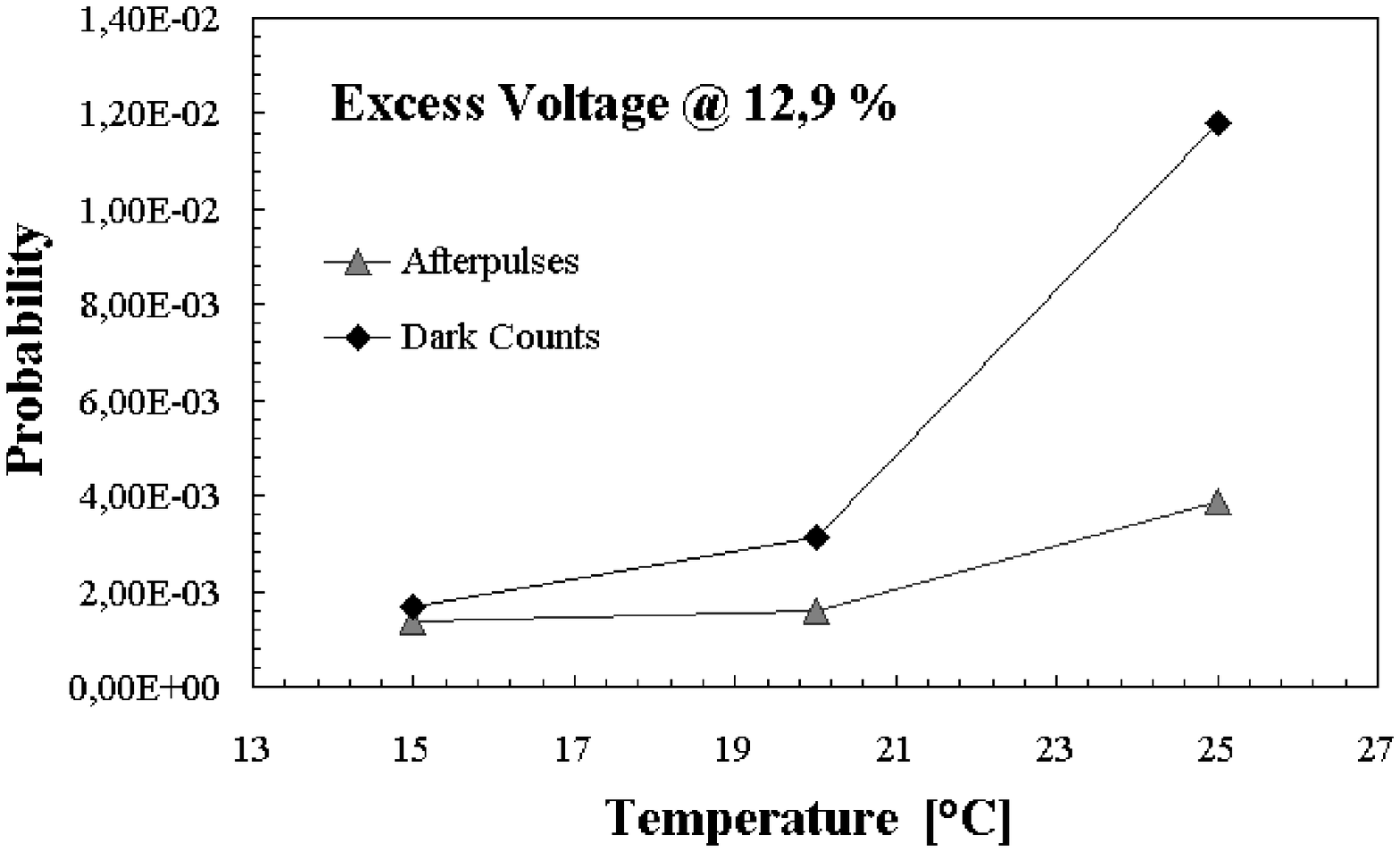,width=11cm,angle=0}
\end{center}
\caption{Total afterpulsing probability for event and the dark counting probability (measured on the 10 $\mu$s window) as functions of the temperature, biased at about 13\% of EBV. The threshold discriminator was fixed to 10 mV.
 }
\begin{center}
\epsfig{figure=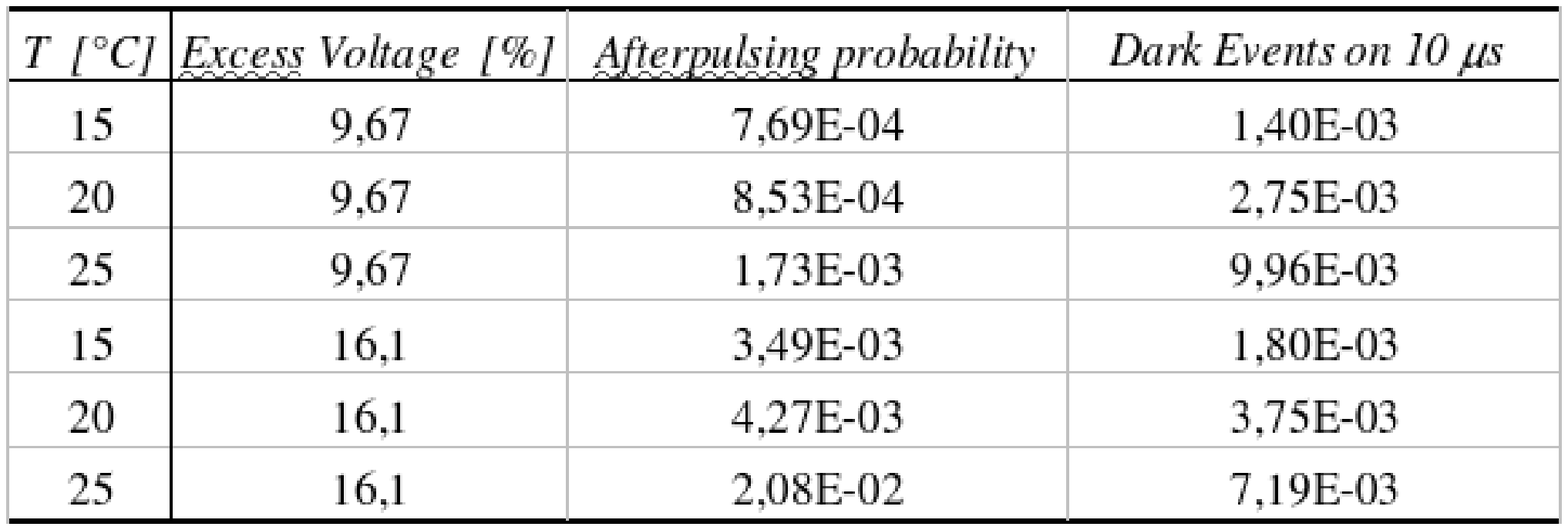,width=11cm,angle=0}
\end{center}
{
Table1. Total afterpulsing probability and the dark counts probability, evaluated by integrating the events in a 10 $\mu$s window, after the subtraction of the uncorrelated background.
 }
\end{figure}
As it arises from figure 7, our experimental data at time $t>t_{max}$ (being tmax the time at which n(t) have the maximum value) are well reproduced by an hyperbolic trend:
\begin{equation}
    n(t)= \frac{n_0}{t^m}
\end{equation}                                                                          
In literature similar trend are reported \cite{musumeci}. Inserting N(t) of the Eq.(2) into p($\gamma$)= $\lambda^{-1}${I(t)} 
follows \cite{abramowitz} p($\gamma$)=$N_0 \gamma^{m-1}$.
By using the relationship between $\tau$ and $\gamma$ it is possible to extract also the P($\tau$)= $N_0/2 \tau^{m+1}$. 
So, this procedure let to describe the kinetics of the charge trapping phenomenon in terms of a probability distribution of the decay constants of the individual process. These distributions have been reported in figure 9 as a function of the temperature. It is important to underline the lack of information around the region of small $\tau$ values as a consequence the physical cut-off on the experimental spectra due to the quenching and successive recharging phenomena.
From such analysis we can conclude that, in the investigated range, the decrement of temperature is correlated with the reduction of  population of the trap levels with high life time values.    
Total afterpulsing probability for event and the dark counting probability (measured on the 10 $\mu$s window) as functions of the temperature, biased at about 13\% of EBV. The threshold discriminator was fixed to 10 mV.
\begin{figure}
\begin{center}
\epsfig{figure=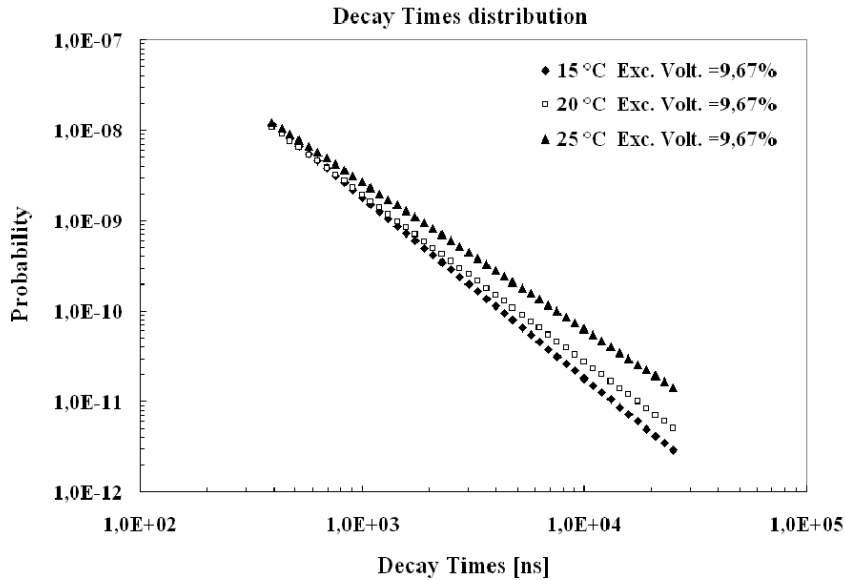,width=12cm,angle=0}
\end{center}
\caption{ Probability function of the trap levels versus decay times, for different temperature and in the particular case of about 10\% of EBV.
 }
\end{figure}

\section{Timing performances }

Due to working principle, avalanche photo-diodes usually provide an excellent timing performance (few hundreds of ps). Such excellent characteristic often are extremely exalted by using ad hoc special fast quenching electronics \cite{cova2}. As far as the first generation devices, tested by using a simple AQC \cite{cova2}, SPADs have demonstrated such excellent performances. A future goal of the next generation of SPAD devices should be the large scale integration, so a good compromise between timing performance and the needed simplicity of the used quenching circuitry on board is required. In this perspective some measurements devoted to the passive quenching technique, in different physical conditions, have been performed \cite{tudisco1,privitera}. 
\\
\begin{figure}
\begin{center}
\epsfig{figure=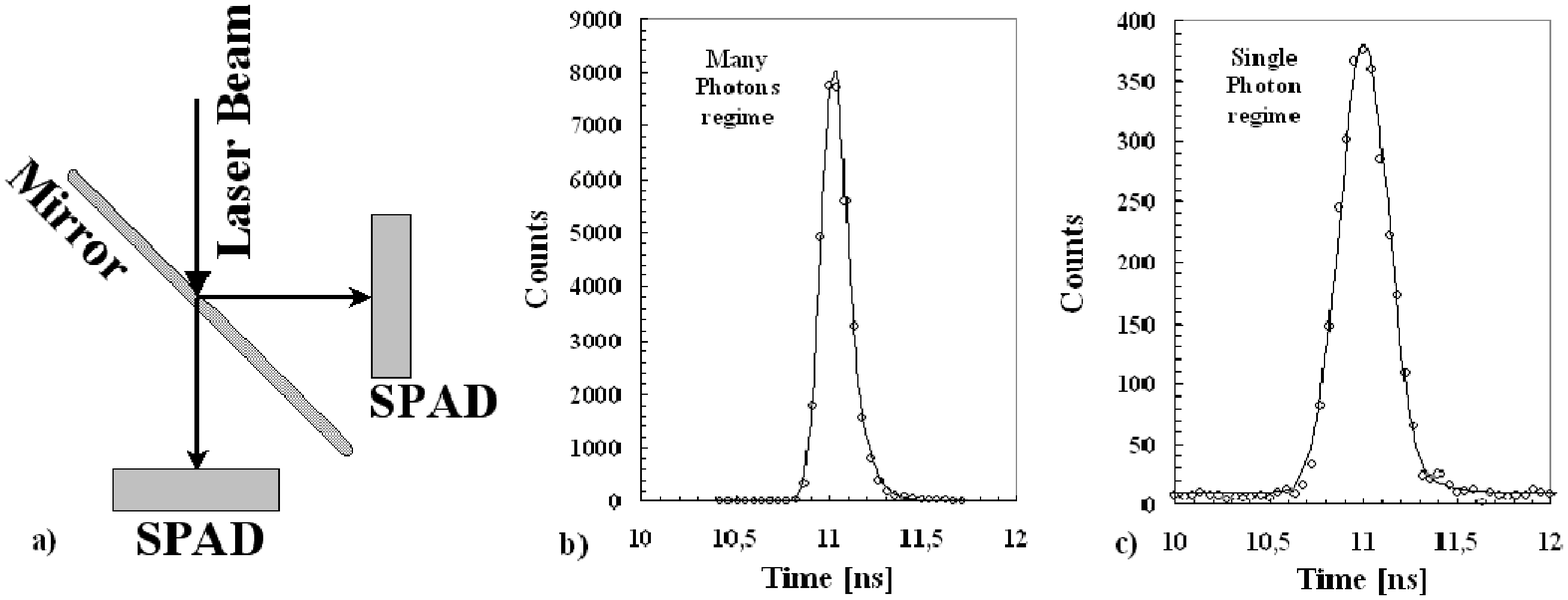,width=14cm,angle=0}
\end{center}
\caption{a) Experimental set-up; b) time spectrum of SPAD with PQC in many photons regime, FWHM $\sim$ 0.161 ns; c) time spectrum of SPAD with PQC in single photon regime, FWHM $\sim$ 0.31  ns
 }
\begin{center}
\epsfig{figure=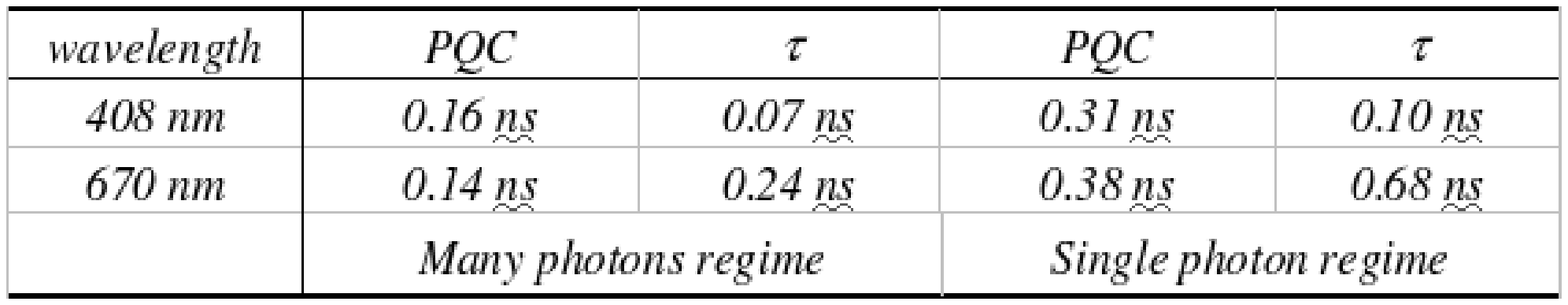,width=11cm,angle=0}
\end{center}
{
Table2. The obtained resulting SPAD time resolutions for the two regime and illuminating conditions
 }
\end{figure}
The employed experimental set-up is reported in figure 5a: an optical pulse from a pico-second laser with two wavelengths 408 nm and 670  nm, used at 1 MHz of maximum repetition rate, was sent, via a semi-reflecting mirror, on two identical SPADs (40 $\mu$m active diameter and at 15\% of EBV). The transition from the many to the single photon regimes was achieved by using a gray filter (with a transmission coefficient of 0.01 \%). A simple electronic chain, based on: a linear Fan-in Fan-out (to reverse the signal polarity) a CFD (Constant Fraction Discriminator), delay module and a TDC (Time Digital Converter) has been used. The TDC start was taken from the laser trigger-out and the individual stops from the signals of each detector.
The transition between many and single photon regime was guaranteed by checking the coincidences between the two SPADs: the probability to have photons in both SPADs, from the same event, was very low (<10-3) and from considerations about the symmetry of the experimental set-up  the probability to have two photons on each detector was negligible.
The results are synthesized in figure 10 and table 2. The excellent time resolution of the SPAD was deduced also in this simple conditions. Timing spectra of the SPADs obtained in both the single an many photons regime are well fitted by a gaussian function plus an exponential tail. The tail depends on the wavelengths because on the penetration depth of impinging photons \cite{cova1}. It is important to remind that timing distribution from measurement in single photon regime include also the timing structure of the laser pulse which we have estimated in 260 ps as FWHM. 

\subsection{Timing profile measurement of a dye laser with SPAD}

In order to prove the feasibility of SPAD sensors as photodetectors for various applications (timing, photon counting etc), also for those where a photomultiplier (PMT) is commonly used, a measurement of the timing structure of a pulsed laser had been performed. The system was constituted of: a UV laser (337 nm wavelength), pulsed with a declared timing resolution of 2 ns (FWHM) and  used with a repetition rate of 30 Hz, which was coupled with a dye laser system for the wavelength shift down to 395 nm. 
To validate the effectiveness of our detection systems the measures have been executed in both many and single photon regimes. The used experimental setup is showed in figure 10: two identical SPADs, with 40 m active diameter, were coupled to the two opposite faces of a trapezoidal shaped piece of Plexiglas. As reference standard readout detector, an HAMAMATSU R6427 20 mm diameter PMT was coupled. We stress that the optical coupling of the photosensors was simply done by putting them in contact with the Plexiglas, as we were not aiming the optimization of the yield. On the contrary, in order to operate also in the single photon regime, the main goal was to decrease the light collection; some gray filters were used. The geometrical efficiency of each SPAD with respect to the PMT was of the order of 10-5. The electronic chain used was the same previously mentioned.  
\begin{figure}
\begin{center}
\epsfig{figure=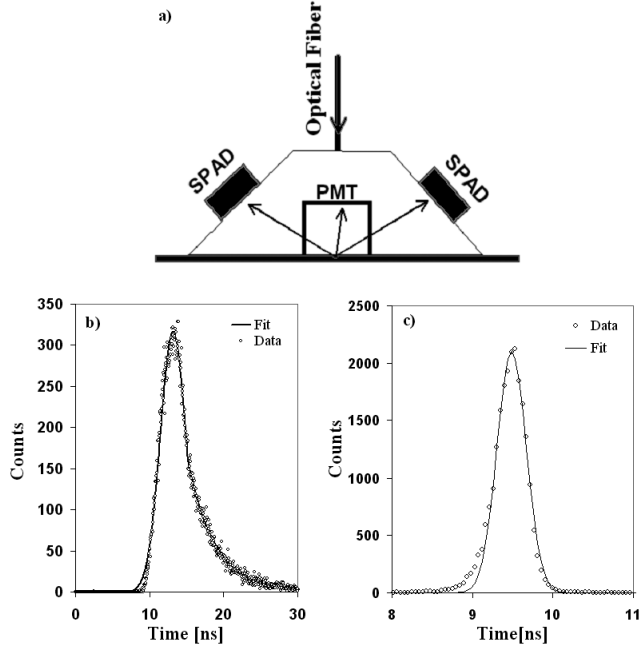,width=12cm,angle=0}
\end{center}
\caption{ a) First experimental set-up; timing spectrum of passively quenched SPAD in: b) many photons regime shows a FWHM $\sim$ 0.161 ns; c) single photon regime shows a FWHM $\sim$ 0.31  ns
 }
\end{figure}
\\
In order to assure the SPADs effective operating conditions, the single photon regime was checked by the evaluation that in no case, within our statistics, the events corresponding to the coincidence of the two SPADs with the PMT (which is firmed by the presence of laser trigger out signal) was present.
In figure 11 the timing spectra of one SPAD in the different regimes, selected by using different filters, are reported. The timing profile for SPAD in the single photon regime is perfectly reproduced by a gaussian plus exponential fit, reflecting the timing structure of the laser pulse, due to the dye excitation, and confirming the single photon operation mode. When detectors operate in many photons regime the obtained resolution is the sum of the two main contributions coming from both the detectors resolution and the laser timing resolution.

\section{5x5 array characterization}

A photomultiplier based on silicon technology represents the new frontier for the photodetection. The integration of SPAD devices on the same substrate, with parallel read-out, makes their combination able to detect the photon arrival position and gives an output pulse directly proportional to the intensity of the source (like a photomultiplier), that is excluded for the single device operating in Geiger-mode.
\\
By using Metal-Resistor-Semiconductor structure is possible to produce SPAD devices together with their integrated quenching circuitry; resistive elements are chosen and embedded for each individual microcell, providing the effective feedback for stabilization and quenching of the avalanche process. With this aim, a first prototype of SPAD arrays have been designed and manufactured \cite{saveliev}. The array is the result of the integration of 25 identical SPAD-STM devices, in the square geometry of 5x5, as shown in the lay-out of figure 2; devices chosen for the integration have the active junction with diameters of 20, 40, 50 and 60 m. Relatively to the different areas, distances between center to center adjacent elements range from 160 to 240 m.  As the picture, cathodes (tagged “C\#”)  are separately contacted, standing available outside the wafer, and all the five anode contacts of each array are connected giving common rows, tagged “A\#”, for the readout of the signal.
Differently by the single element SPAD, in case of the array configuration the local gettering region uniformly surrounds the active area of each pixel through an external ring doped by heavy phosphorus diffusion, which also provides the decrease in the dark counting rate. In order to reach the uniformity of the breakdown within the entire active area, a virtual guard ring using a large window of n+ polysilicon is created; moreover, a good uniformity (less then 1\%) on the whole devices of the array had been founded. A sampling on dark counting rates on 30 equal arrays of 5x5 equal elements with an active area of 20 m in diameter have been showed a very narrow statistical distribution with an average value of about 400 cps and a dispersion of 50 cps \cite{sciacca3}.
\\
Other limitations to the photon counting arise from the integration of adjacent devices: by the ignition of the avalanche process in a SPAD, spurious uncorrelated avalanches may be triggered in the neighboring devices, due to the both possible effects of the electromagnetic and/or optical induction. This effect, which is called cross-talk, produces spurious pulses, so, increasing the detector noise. In fact, is an evidence that, when reversely biased over the breakdown value, a silicon p-n junction diode emits photons; the emission probability was estimated about 10-5 photons per carrier crossing the junction \cite{lacaita2}. So, some avalanche process may be originated in the near detector. Here, the deep investigation and study about both the contributions to the cross-talk effect, occurring among SPAD-STM devices of a 2-D array, will be presented.
Optical cross-talk takes place when the hot-carriers, generated during the avalanche process, emit secondary photons by radiative emissions \cite{lacaita2}. In silicon, a photon may travel with an attenuation length of 80 µm, for photons in the near UV and visible region; this represents the fast component of the cross-talk. When produced, these photons can be absorbed in the same junction or on the closer SPAD devices, triggering an avalanche process. Such optical cross-talk may be minimized by both a suitable optical isolation among the diodes (if the pixels are very closed) and reducing the number of carriers during the avalanche process.
\\
On the other hand, the electromagnetic cross-talk, occurs when hot-carriers, generated during the avalanche process, overcame the junction, reaching the neighbor device and triggering an avalanche multiplication process. Carrier velocity depends on the electric field strength. Typical value of mobility for electrons in silicon at room temperature is $\mu_e$=1500 $cm^2 / V^*$ sec; so, the velocity of the overcoming electron arises from the product of the mobility and the field strength outside the junction, which is less than the 105 V/cm. Because that velocity is many orders of magnitude lower than that of the photon, it represents the slow component of the cross-talk; it is an important parameter to calculate the effective electrical contribution among detectors located on the same substrate and it can be waited for its dependence on the distance between elements.
\\
Another electromagnetic contribution is given by the electric connections of the pixels, which may became an important effect when a high density of elements are implemented.
In order to avoid the optical cross-talk, during the fabrication of arrays of SPAD-STM a delicate process connects “trenches” with metal coated sidewalls (into the bulk of semiconductor) between pixels, reducing the minimum distance between elements and increasing the dynamic range of the device. This is a delicate process because metal is posed close to the pixel after a difficult previous removing from the active region. Actually, arrays of SPAD-STM optically and electrically isolated by deep thin trench technology was designed and fabricated (see section 2). 
The investigation was started with the analysis, to the oscilloscope, of the dark signals coming from elements of an array of 5x5 passively quenched SPAD-STM with an active diameter of 20 m, as from the lay-out of figure 2. Firstly, it can be observed the highly dense region, where the anodic rows are very closed to each others.
The possible electric and electromagnetic contribution between the two detectors was tested: each time any pixel signal is chosen as trigger (by a dark event) the contributions induced on both the neighbor and far rows were observed. 
The results from the samples with and without trench showed the absence of any effect optical effect induction related to the inter-pixels distance, as expected from the high pitch values. Moreover, the comparison between the signals shows the presence of small pulse (its amplitude is about factor ten smaller) of opposite polarity in with respect to the signal trigger and with a time structure very close to the its generator. Such effect probably come from an electric contribution between pixels, due to the resistive nature of the substrate, this is the unique way able to determine an alteration of the field conditions for each detector inside the structure.
\\
After this preliminary analysis a much more deep study of the cross-talk effect has been done by means of the time correlation measurements on a couple of elements of the array. We arranged an experimental set-up based on a 32 channel multi-hit Time to Digital Converter (TDC) module, mounted on VME bus and a data acquisition system realized with standard nuclear electronics. The input signals for the TDC, was taken from couples of elements of the 20 m array. The system is able to correlate signals in 50 µs starting from a master trigger; here this signal was generated by dark events of one of the two detectors. Moreover, we investigated both the two arrays with and without the optical isolation trenches between individual channels. We found no difference between the two cases: evidence that in such device the optical cross-talk is negligible. A preliminary test with two detectors of two different arrays had been showed two completely uncorrelated spectra, so tested the right operating conditions of the set-up. The signal of the detector which acts as trigger was send on one channel (Ch1) and taken from the element 5-1 of the matrix; in order to investigate the inter-pixel distance dependence, the signal on the other channel (Ch2) was delayed and alternatively chosen between the detectors 4-1, 4-2, 3-1 and 1-1. In a such way data from the Ch1 are the number of start events inclusive of the contribution of delayed pulses; data on Ch2 are the cross correlated effects.
\\
\begin{figure}
\begin{center}
\epsfig{figure=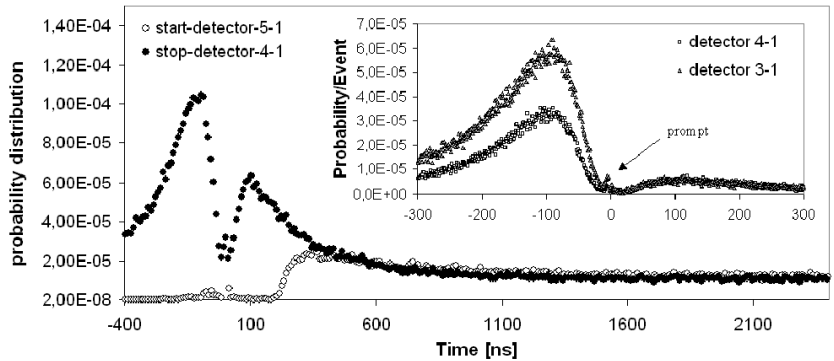,width=12cm,angle=0}
\end{center}
\caption{ Spectra of the Ch1 (pixels 5-1) and Ch2 (4-1) of TDC normalized to the total number of start.
 }
\end{figure}
In figure 12 was reported Ch1 end Ch2 spectra (normalized to the total number of start) related to the measure of the pixels 5-1 (start detector) and 4-1; as expected from the previous observation, discussed in the section 4.2, the Ch1 spectrum showed both the two contributions of the uncorrelated dark events and the after pulses modulated in the first hundreds of nanoseconds from the quenching mechanism and the successive recharging, on which acts the discriminator threshold. 
A similar structure, on a uncorrelated dark background of the 4-1 pixel, was observed on both the two temporal regions of Ch2 spectrum; before and after the time zero which corresponds to the trigger signal coming from the pixel 5-1. The correlated events before the signal trigger are connected to the primary avalanches generated inside pixel 4-1 and the successive correlation to the other pixel 5-1. On the contrary, all events succeeding the (time) zero are connected to the primary avalanches generated inside the pixel 5-1, which trigger the acquisition, and the successive correlation to the signal from detector 4-1. The strength  of such contributions seems to be quite similar to that one observed for the correlated after pulses showed on Ch1 spectrum, but this case without any affect due to the quenching mechanism and successive recharging phase. Following this indication, it was decided to perform, on data from both the two channels, a similar analysis procedure as it have been discussed in section 4.2. Some of the obtained results of such procedure had been reported in figure 13, where the discovered difference between timing behavior is showed; in particular, also in this case the after-pulses distribution, related to the detector 5-1, used as start detector in the whole set of measurements, follows the behavior showed in section 4.2. 
\\
A faster kinetics seems to characterize the correlated events distribution, without any relevant difference in respect to the inter-pixel distance. We interpreted such observations as a probable phenomenon of charge trapping and successive releasing from the no active zones surrounding the pixels, the common substrate. 
Moreover, the cross correlated events reported in the insert of figure 12 shows the prompt contribution, as from precedent measurements \cite{finocchiaro}, confined in the first nanoseconds and increasing with the excess bias voltage.
In the framework of our interpretations such phenomenon is due to an effect of the electric contribution between pixels, connected to the resistive nature of the substrate: a pixel breakdown determines an alteration of the field conditions on the other detector inside the structure, which acts for instance on the bias voltage across the neighbor element and induces its avalanche.  
\begin{figure}
\begin{center}
\epsfig{figure=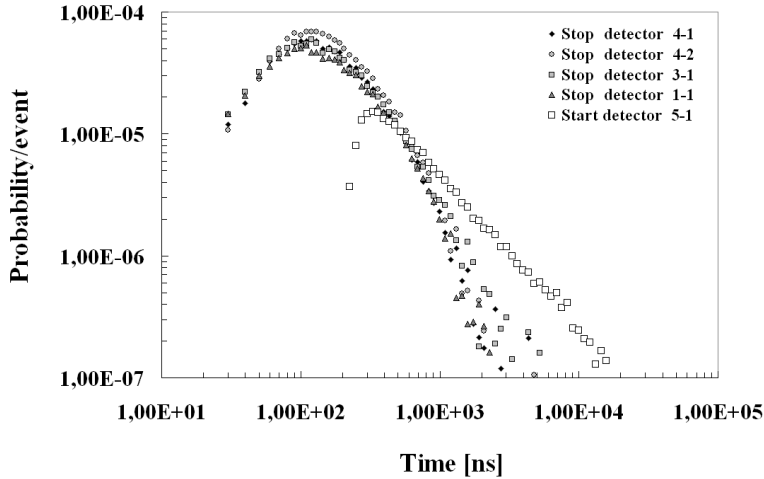,width=12cm,angle=0}
\end{center}
\caption{ Timing distribution of the events related to the start (afterpulsing) and stop detectors (cross talk).
}
\end{figure}

\section{SiPm concept}

One of the major goal in the field of photonics is the realization of a new concept photo-detector like a photomutiplier tube (PMT) and with all the potential advantages of the silicon based technology. Such device can be realized from the common readout of bi-dimensional arrays of SPAD detectors \cite{saveliev,buzhan,campisi,dolgoshein}. The so-called SiPM should provide: high sensitivity to the single photon counting, high quantum detection efficiency over a wide part of the spectrum (up to the near IR region), insensitivity to the magnetic field and very low bias voltage.
\\
\begin{figure}
\begin{center}
\epsfig{figure=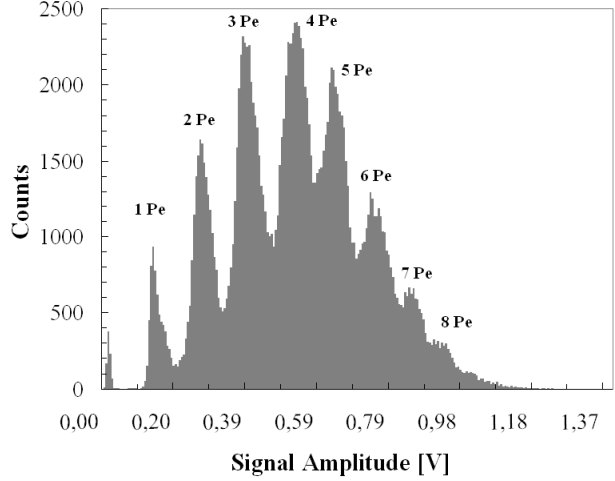,width=12cm,angle=0}
\end{center}
\caption{ Figure 14. The amplitude distribution of measured signals from the 5x5 array of SPAD devices of 20 $\mu$m on active diameter, obtained illuminating by a laser with a $\lambda$=670 nm; the EBV equals 10\% and at the steady temperature is 20°C.
 }
\end{figure}
In this section, a first prototype of a 5x5 array of SPAD devices manufactured in ST-Microelectronics of Catania, its preliminary characterization in SiPM configuration will be presented. 
SiPM configuration results from the parallel readout of every SPAD element, each one of which are passively quenched  by means of a ballast resistor of 100 k$\Omega$; output signal is detected on the common load resistor RL which connect all the anodes to the ground. In this architecture, the signal is the sum of all the individual cells fired by the photon-initiated avalanche phenomenon. Each single SPAD element operates as a binary device, while their combination makes the device an analogue detector, like a proportional counter, able to give information about the intensity of the illuminating source. The output current signal observed to an oscilloscope manifests a multiple structure with several amplitudes: for example, when two photons are simultaneously detected by two (different) pixels a signal with double amplitude was expected; and so on, when n photons are simultaneously detected by n different pixels the total output signal was expected an amplitude n-times that one of the single. Noise and cross-talk effects limit the realization of highly dense structure: typical density of cells may extend up to few thousands per mm2, limiting the dynamic range of the device. 
\\
\begin{figure}
\begin{center}
\epsfig{figure=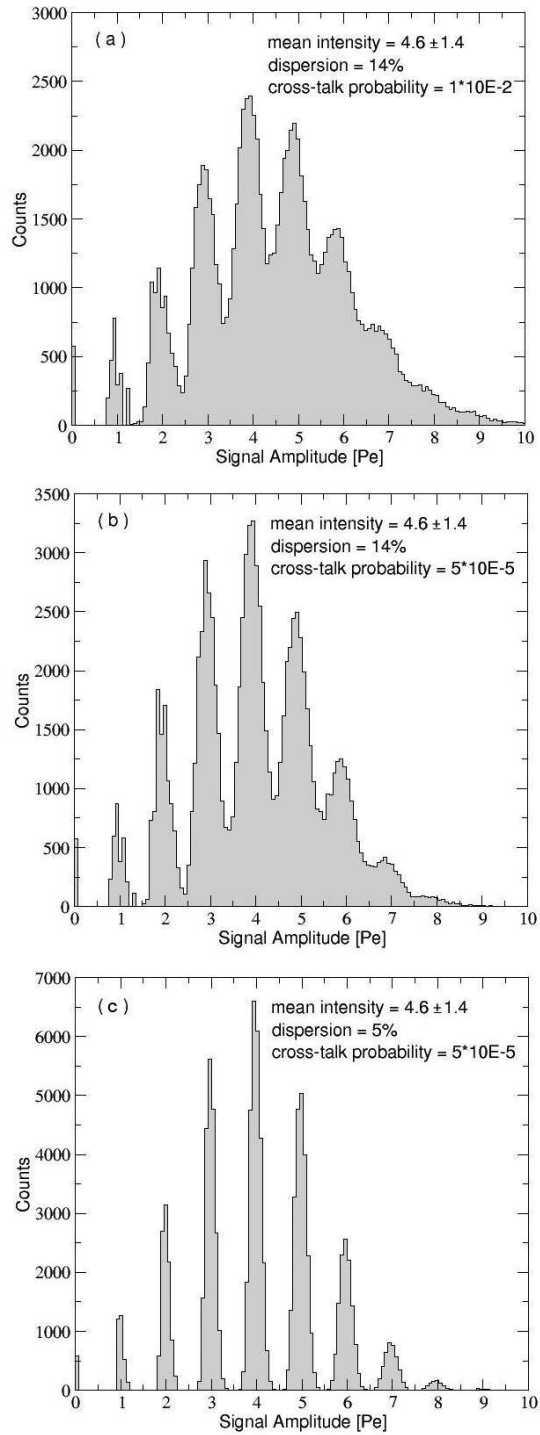,width=8cm,angle=0}
\end{center}
\caption{ Montecarlo simulated distributions of the signal amplitude from a 5x5 SPAD array in a SiPM configuration once illuminated; comparison between three different sets of input parameters.
}
\end{figure}
As previously mentioned, the present characterization had been centered on 5x5 array of devices with an active area of 20 $\mu$m on diameter and a pitch of 200 $\mu$m, optically isolated by means of an opaque trench. In order to architecture SiPM configuration and provide the resulting signal, all the five anodic rows are connected each other. To test the response as a function of the number of impinging photons, the array was illuminated by picosecond laser of adjustable in intensity and with a $\lambda$=670 nm, used at a repetition rate of 10 kHz. The output signal had been converted by using an amplitude to digital converter (ADC). The array had been biased at 10\% of the EBV and cooled at the steady temperature of 20 °C. The amplitude distribution of measured signals has been reported in figure 14. The single (double, triple, etc.) photoelectron peaks(s) was clearly visible, demonstrating, also in this hybrid configuration, the pixels excellent performance and the good single photoelectron resolution of the device.  
In order to perform a deep investigation on the effects of the integration of many elements on the spectrum response on figure 14, the detectors response and the cross-talk (as discussed up to this moment) have been simulated by an ad-hoc Montecarlo code. 
Such code simulates the amplitude distribution of signals from an array illuminated by a laser pulse, starting from the single pixel response in terms of profile, total duration, rise and fall time. The input parameter of the code was:  
 (i) the laser intensity and its dispersion, representing the number of fired pixels normally distributed (ii) the percentage of non-uniformity among all the pixels, essentially due to the hybrid configuration; external circuitry, solders, etc (iii) the cross-talk probability, assumed with an infinite interaction range (iiii) the dark counting rate.
The input parameters have been fixed to those quantities obtained by the experimental (presented) results. In particular, a comparison between the effects of non-uniformity on the single element response and cross-talk by means of the observation of the distribution, had done.
The results of such procedure have been reported in figure 15b; the extracted 14\% of non-uniformity parameter was in reasonable agreement with the experimental observation. It appears that the increasing on the non-uniformity generates an asymmetric distortion of the amplitude distribution and the cross-talk acts on the Gaussian-like background (see figure 15a) shifting the distribution towards high photoelectron picks. Particularly, from the comparison between the figures 15b and 15c, arising from the same value of the cross-talk probability, the reduction of the non-uniformity on the detector response makes the spectrum very well resolved.

\section{Conclusions}

In this work the full characterization of the single photon avalanche diode, SPAD, manufactured by ST-microelectronis and passively quenched is presented. The really promising results in terms of PDE, dark counting rate, timing, after pulsing probability demonstrate that SPAD device is an ideal candidate among the existing single photon sensors.
The integration possibility has been also investigated by 5x5 arrays manufacture. Such devices have been tested in order to study the dark counting rate uniformity, the cross-talk phenomenon and to perform the preliminary test on the SiPM configuration. The substrate contribution seems to be the most probable candidate in order to explain the results of cross-talk measurements. Moreover, as evident from the simulation, a high uniformity level between elements is the major contribution on the quality of the SiPM response (spectrum).

\end{document}